\documentclass[12pt]{iopart}
\usepackage{graphicx}
\usepackage{color}
\begin{document}

\title{Torsion-induced persistent current in a twisted quantum ring}

\author{Hisao Taira$^{\rm 1)}$ and Hiroyuki Shima$^{\rm 1,2)}$}

\address{${}^{\rm 1)}$ Department of Applied Physics, Graduate School of 
Engineering, Hokkaido University, Sapporo 060-8628, Japan }

\address{${}^{\rm 2)}$ Department of Applied Mathematics 3, 
LaC${\rm \grave{a}}$N, Universitat Polit${\rm \grave{e}}$cnica de Catalunya 
(UPC), E-08034 Barcelona, Spain}

\ead{taira@eng.hokudai.ac.jp}
\begin{abstract}
We describe the effects of geometric torsion on the coherent motion of 
electrons along a thin twisted quantum ring. 
The geometric torsion inherent in the quantum ring 
triggers a quantum phase shift in the 
electrons' eigenstates, thereby
resulting in a torsion-induced persistent current that flows along 
the twisted quantum ring. 
The physical conditions required for detecting 
the current flow are discussed.
\end{abstract}

\pacs{73.23.Ra, 73.21.Hb, 02.40.-k, 03.65.Ca}
\maketitle

\section{INTRODUCTION}

Spatial confinement of particle's motion to low-dimensional space has an 
enormous influence on the quantum-mechanical properties of 
the particle. 
Of particular interests are systems in which a
particle's motion is constrained to a thin curved layer by a strong 
confining force. 
Due to the confinement, excitation energies of the 
particle in a direction normal to the layer are significantly higher 
than those in a direction tangential to it; as a result, one can
define an effective 
Hamiltonian that involves an anisotropic effective mass and a 
curvature-induced scalar potential
\cite{math1,math2,maraner}. 
This implies that the behavior of quantum particles that are confined to
a thin curved layer is different from that of quantum particles on
a flat plane, even in the absence of external field 
(except for the confining force). 
The effect of curvature was first suggested by Jensen and Koppe \cite{math1}, 
and this was followed by subsequent studies that were conducted 
out of mathematical curiosity \cite{math6}. 
In recent years, the effect of curvature has been reconsidered 
from the viewpoints of condensed matter physics \cite{phys2,phys3,
phys8,phys9,encinosa,phys10,phys12,phys13,phys14,
phys15,phys16}, 
owing to technological progress that has enabled the fabrication of  
nanostructures with curved geometries
\cite{exp1,exp2,exp4,exp6,jpcm,exp8}.

In addition to surface curvature, geometric {\it torsion} is another 
important parameter relevant to quantum mechanics in low-dimensional
nanostructures. 
A torsion effect is manifested in quantum transport in a thin twisted 
nanowire with a finite cross section. 
When a quantum particle moves along a long thin twisted wire, it exhibits
a quantum phase shift whose magnitude is 
proportional to the integral of the torsion along the wire
\cite{torsion1,torsion2}. 
This torsion-induced phase shift is attributed to an effective 
vector potential that appears in the effective 
Hamiltonian defined for the movement of a particle in a twisted 
nanowire. 

The mathematical mechanism for the occurrence of the effective 
vector potential was demonstrated by Takagi and Tanzawa \cite{torsion1},
and independently by Magarill and ${\rm \acute{E}}$ntin \cite{entin}. 
Their results imply various intriguing phenomena 
purely originating from geometric torsion.
For instance, the torsion-induced phase shift may give rise to a novel class of
persistent current flow along a closed loop of a twisted wire;
it is novel in the sense that no magnetic field need to penetrate inside the loop, which is in contrast with the ordinary persistent current \cite{persistent1,persistent2,persistent3,persistent4,persistent5,persistent6,persistent7,persistentderivation2} observed in a non-twist quantum loop.
However, optimal physical conditions as well as geometric parameters
in order to measure those phenomena have been overlooked so far.
Quantitative discussions as to what degree of torsion is necessary to make the phenomena
be measurable in real experiments are important from both fundamental and practical 
viewpoints. 

In this article, we have investigated the quantum state of electrons 
in a closed loop of a twisted wire, i.e. a twisted quantum ring.
The wire consists of twisted atomic configuration, and its centroidal 
axis is embedded in a flat plane; 
these assumptions mean that the torsion in our system is defined with 
respect to a twisting crystalline reference frame.
We have revealed that the magnitude of the torsion-induced persistent 
current $I$ comes within a range of existing
measurement techniques under appropriate conditions; 
this indicates the significance of torsion-induced 
quantum phase shift in the study of actual 
nanostructures, besides its theoretical interest. 
It should be emphasized that the persistent current $I$ we have considered
is free from a magnetic field penetrating 
through the ring, and thus differs inherently from the counterpart observed
in untwisted rings. 

\section{QUANTUM STATE IN A TWISTED WIRE}

\begin{figure}[t]
\begin{center}
\includegraphics[width=8cm,height=3cm]{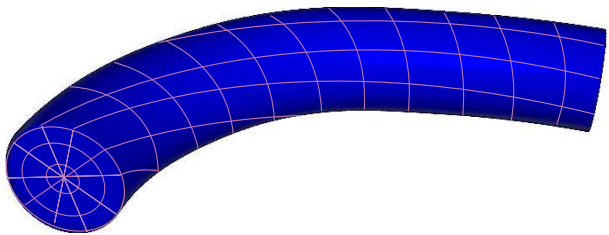}
\caption{\label{fig:fig1} 
Sketch of a twisted quasi-one
dimensional wire with a circular cross section. 
The mesh indicates the curvilinear coordinate $(q_0,q_1,q_2)$ 
used in this study.
Geometric torsion of the atomic configuration along the 
cylindrical axis is represented by the rotation of the reference
frame in cross section (see text).}
\end{center}
\end{figure}

In this section, we derive an explicit form of the effective vector potential 
in line with the discussions presented in reference \cite{torsion1}.
Let us consider an electron propagating in a long thin curved cylinder 
with a weakly twisted atomic configuration (figure \ref{fig:fig1}). 
For simplicity, 
the cylinder is assumed to have a circular cross section with constant
diameter $d$. 
We introduce orthogonal curvilinear coordinates 
$(q_0, q_1, q_2)$
such that $q_0$ parametrizes the centroidal axis $C$ of the curved 
cylinder (i.e. the curve $q_1=q_2=0$ coincides with $C$).
We assume that $C$ is embedded in a flat
plane so that $C$ {\it itself} has no torsion; therefore, 
the torsion of the present system
is a consequence of the twisted atomic structure around the axis $C$ of 
the conducting cylinder.

A point on $C$ is given by the position vector 
$\bi{r} \equiv \bi{r}(q_0)$. Similarly, a point in the vicinity of 
$C$ is represented by
\begin{eqnarray}
\bi{R}=\bi{r}(q_0)+q_1\bi{e}_1(q_0)+q_2\bi{e}_2(q_0),
\label{eq:positionvector}
\end{eqnarray}
where the set ($\mbox{\boldmath $e$}_0,\mbox{\boldmath $e$}_1,
\mbox{\boldmath $e$}_2$) with $\bi{e}_0 \equiv \partial_0 
\bi{R}$ and $|\bi{e}_1|=|\bi{e}_2|=1$
forms a right-handed orthogonal triad; 
we use the notation 
$\partial_a \equiv \partial/\partial q_a \hspace{1.0mm} (a=0,1,2)$ 
throughout the paper. 
Here, the unit vectors $\bi{e}_1$ and $\bi{e}_2$ span the cross section normal to $C$,
and they rotate along $C$ with the same rotation rate as that of the atomic configuration.
To be precise, the $q_0$-dependences of 
$\mbox{\boldmath $e$}_1$ and $\mbox{\boldmath $e$}_2$ are
chosen such that the torsion $\tau$ defined by 
\begin{eqnarray}
\tau = \mbox{\boldmath $e$}_2 \cdot 
\partial_0 \mbox{\boldmath $e$}_1
\label{eq:torsion}
\end{eqnarray}
conforms to that of the twisted atomic structure.
Using continuum approximation, we obtain
the Schr\"odinger equation for the twisted quantum cylinder as
\begin{eqnarray}
-\frac{\hbar^2}{2m^*}\sum_{a,b=0}^{2}\frac{1}{\sqrt{g}}\partial_a
\left(\sqrt{g}g^{ab}\partial_b \right)\phi
+V\phi=E\phi.
\label{eq:syure1}
\end{eqnarray}
Here, $m^*$ is the effective mass of the electron, and 
$V=V(q)$ with $q \equiv \left(q_1^2 + q_2^2\right)^{1/2}$ 
is a strong confining potential that confines the electron's motion 
to the vicinity of $C$.
$g^{ab}$ are elements of the matrix $[g^{ab}]$, which is the inverse of 
$\left[ g_{ab} \right]$ whose elements are
$g_{ab} = \partial_a \mbox{\boldmath $R$} \cdot \partial_b 
\mbox{\boldmath $R$}$ and $g = \det [g_{ab}]$ \cite{shima}.
From equation(\ref{eq:positionvector}), we obtain the following 
explicit forms of $g^{ab}$:
\begin{eqnarray}
g^{00}&=&\gamma^{-4}, \hspace{1.0mm} g^{0a}=\gamma^{-4}
\tau\epsilon_{0ab}q_b, \nonumber \\
g^{ab}&=&\delta_{ab}+\gamma^{-4}\tau^2 
\left( |q|^2\delta_{ab}-q_a q_b \right),
\hspace{1.0mm} [a,b=1,2]
\label{eq:metric}
\end{eqnarray}
where $\gamma=(1-\kappa_a q_a)^{1/2}$ and 
$\kappa_a = \mbox{\boldmath $e$}_0 \cdot \partial_0 \mbox{\boldmath $e$}_a$; 
the summation convention was used in equation (\ref{eq:metric}). 
The quantity $\kappa\equiv(\kappa_1^2+\kappa_2^2)^{1/2}$
represents the local curvature of $C$. 
Note that both $\tau$ and $\kappa$ are functions only of $q_0$.
 
Hereafter, we assume that the geometric 
modulation of the cylinder (i.e. torsion and curvature) is 
sufficiently smooth and small so that the relations 
$\kappa d \ll 1$ and $\tau d \ll 1$ are satisfied. Under these conditions,
equation(\ref{eq:syure1}) is reduced to \cite{torsion1}
\begin{eqnarray}
\mu \left[\left( \partial_1^2 + \partial_2^2 \right) 
+ \left(\partial_0 - \frac{i \tau L}{\hbar} \right)^2
+\frac{\kappa^2}{4} \right]\phi +V \phi = E \phi,
\label{eq:syure2}
\end{eqnarray}
where $\mu \equiv -\hbar^2/(2m^*)$ and 
$L\equiv -i\hbar(q_1 \partial_2 - q_2 \partial_1)$
is the angular momentum operator in the cross section.
The solution for equation(\ref{eq:syure2}) is assumed to have the form
\begin{eqnarray}
\phi(q_0, q_1, q_2)=\psi(q_0) \sum^N_{j=1} c_j u_j(q_1, q_2).
\label{eq:separationvariables}
\end{eqnarray}
Here $u_j(q_1,q_2)$ is an $N$-fold degenerate eigenfunction of 
the operator of 
$H_{\bot} \equiv \mu \left( \partial_1^2 + \partial_2^2 \right) +V(q)$
that is invariant to the rotation of the coordinates $q_1$, $q_2$. 
This means that $u_j(q_1,q_2)$ is an eigenfunction of $L$ such that
\begin{eqnarray}
Lu_j(q_1,q_2)=\hbar m_j u_j(q_1,q_2),
\label{eq:angulareigenfunction}
\end{eqnarray}
where $m_j$ is an integer. 
Thus, we multiply both sides of 
equation(\ref{eq:syure2}) with $\sum_j c_j^* u_j^*(q_1,q_2)$ 
and integrate with respect to $q_1$ and $q_2$ in order to obtain an 
effective one-dimensional equation,
\begin{eqnarray}
\mu \left[\left( \partial_0- \frac{i\tau \langle L \rangle}{\hbar} \right)^2
+\frac{\kappa^2}{4} - \frac{\tau^2}{\hbar^2} \left( \langle L^2 \rangle - \langle L \rangle^2 \right)\right]\psi(q_0) = \epsilon \psi(q_0),
\label{eq:1dsyure2}
\end{eqnarray}
where $\langle L \rangle = \hbar \sum_j |c_j|^2 m_j$ and $\epsilon$ 
is the eigenenergy of an electron moving in 
the axial direction. 
The product $\tau \langle L \rangle$ in 
parentheses is identified to the effective vector potential
 mentioned earlier.
\section{TORSION-INDUCED PERSISTENT CURRENT}
\begin{figure}[t]
\begin{center}
\includegraphics[width=6.0cm]{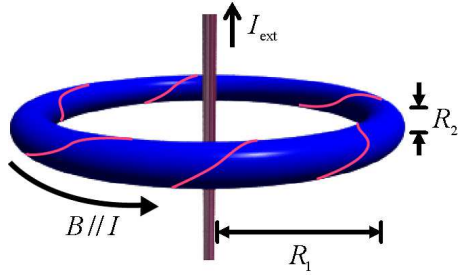}
\caption{\label{fig:fig3} 
Twisted quantum ring encircling external 
current flow $I_{\rm ext}$. 
A magnetic field $B$ induced along the ring breaks the 
time reversal symmetry of the system, thus resulting in a torsion-induced 
persistent current $I$ parallel to $B$.}
\end{center}
\end{figure}
We now consider a closed loop of a twisted quantum wire with a circular 
cross section of constant radius $R_2$, which we call a twisted quantum ring. 
For simplicity, the centroidal axis $C$ of the ring
 is set to be a 
circle of radius $R_1 \gg R_2$, which results in a constant curvature 
$\kappa \ll 1/R_2$ (i.e. $q_0$-independent). 
In addition, we assume that the torsion $\tau$ of the atomic 
configuration around $C$ is constant throughout the ring and satisfies
the condition $\tau R_2 \ll 1$ (generalization to the case in which $\kappa$ 
and/or $\tau$ are $q_0$-dependent is straightforward). 
Hence, an electron's motion in the twisted ring is 
described by equation(\ref{eq:1dsyure2}), from which we obtain
\begin{eqnarray}
\psi(q_0) = \psi_{\rm unt}(q_0) \exp\left(-i \frac{\tau}{\hbar} \int_0^{q_0}
\langle L \rangle dq_0'\right), 
\label{eq:solution}
\end{eqnarray}
where $\psi_{\rm unt} \propto \exp(-ikq_0)$ is the eigenfunction 
of an untwisted ring (i.e. $\tau \equiv 0$). 
An additional quantum phase proportional to 
$\tau$ implies the presence of a torsion-induced persistent 
current throughout the ring, as will be proved below.

Equation(\ref{eq:solution}) shows that the condition 
$\langle L \rangle \neq 0$ is necessary for the presence of a 
torsion-induced persistent current. 
The condition can be realized by applying an external current $I_{\rm ext}$
that penetrates through the center of the ring, as shown in 
figure \ref{fig:fig3}.
Using the polar coordinate system ($r,\theta$)
with respect to the circular cross section, $L$ in equation(\ref{eq:syure2}) 
is rewritten as
\begin{eqnarray}
L_B=-i \hbar\frac{\partial}{\partial \theta} -\frac{eBr^2}{2},
\label{eq:angularoperator} 
\end{eqnarray}
where $B=\mu_0 I_{\rm ext}/\ell, \hspace{1.0mm} \ell=2\pi R_1$ and $\mu_0$ 
is the permeability constant. 
The confining potential $V(r)$ 
is set to be a parabolic well
centered at $r=0$, $V(r) = m^* \omega_p^2 r^2 /2$, where
$\omega_p$ characterizes the steepness of the potential. 
Hence, the lowest energy eigenstate $u_0$
in the cross section is given by \cite{Fock,Darwin} 
\begin{eqnarray}
u_0(r)=\sqrt{\frac{m^*\Omega}{\pi \hbar}}\exp\left(-\frac{m^*\Omega}{2\hbar}r^2\right),
\label{eq:u0} 
\end{eqnarray}
where $\Omega=\sqrt{\omega_p^2+(\omega_c/2)^2}$ and $\omega_c=eB/m^*$ is 
the cycrotron frequency.
As a consequence, the expectation value of $L_B$ with respect to
$u_0$ reads
\begin{eqnarray}
\langle L_B \rangle =\displaystyle\int_0^{\infty}rdr\int_0^{2\pi}
d\theta u_0^* L_B u_0
=-\frac{\hbar eB}{2m^* \Omega},
\label{eq:expectation1} 
\end{eqnarray}
or equivalently, 
\begin{eqnarray}
\langle L_B \rangle = -\frac{e\mu_0\hbar}{2\ell m^* \left[\omega_p^2+\left(\frac{e\mu_0}{2\ell m^*}I_{\rm ext}\right)^2\right]^{1/2}}I_{\rm ext}.
\label{eq:expectation} 
\end{eqnarray}
From equation(\ref{eq:expectation}), we see that $\langle L_B \rangle \neq 0$
if $I_{\rm ext} \neq 0$.

The persistent current $I$ driven by $\tau$ is evaluated by considering
the periodic boundary condition $\psi(q_0 + \ell) = \psi (q_0)$
that holds for the twisted ring.
Since $\psi_{\rm unt}(q_0) \propto \exp(-ikq_0)$, it follows from 
equation(\ref{eq:solution}) that
\begin{eqnarray}
\exp(-ik\ell) \exp \left( -\frac{i}{\hbar}\tau\langle L_B \rangle \ell 
\right) =1 ,
\label{eq:equivalently}
\end{eqnarray}
or equivalently,
\begin{eqnarray}
k=\frac{2\pi}{\ell} \alpha - \frac{\tau\langle L_B \rangle}{\hbar} \equiv 
k_{\alpha}, \hspace{2.0mm} (\alpha =0,\pm 1 , \pm 2 \cdots).
\label{eq:wavenumber}
\end{eqnarray}
The current carried by a single electron in the $\alpha$th eigenstate 
is $I_{\alpha} = e v_{\alpha} /\ell=e \hbar k_{\alpha} /(m^*\ell)$
\cite{persistentderivation}.
The total persistent current $I$ in a ring 
containing $N$ electrons at zero temperature is obtained by summing 
the contributions from all eigenstates with energies less than $E_F$. 
It is known that $I$ for odd $N$, denoted by $I_{\rm odd}$, 
differs from that for even $N$, denoted by $I_{\rm even}$
\cite{persistentderivation}\footnote{It is noteworthy that a complete description of the sign and magnitude of the persistent current
for non-twisted rings has been recently proposed in reference \cite{imry}
by considering the role of electron-electron interactions.}. 
In fact, straightforward calculation yields
\begin{eqnarray}
I_{\rm odd}&=&2\times\sum_{\alpha=-(N-1)/2}^{(N-1)/2}I_{\alpha}
=2\times\sum_{\alpha=-(N-1)/2}^{(N-1)/2}\frac{e\hbar}{m^*\ell}
\left( \frac{2\pi}{\ell} \alpha - \frac{\tau\langle L_B \rangle}{\hbar}
\right) \nonumber \\
&=&-\frac{ev_F}{\ell}p, \hspace{0.5cm} {\rm for} \hspace{0.5cm} -2\leq p <2 \\
\hspace{-2.5cm} {\rm and} \nonumber \\
I_{\rm even}&=&2\times\sum_{\alpha=-N/2+1}^{N/2}I_{\alpha}
=\frac{ev_F}{\ell}\left( 2- p \right), 
\hspace{0.5cm} {\rm for} \hspace{0.5cm} 0\leq p <4
\label{eq:persistent_oddandeven}
\end{eqnarray}
where $v_F \equiv \pi \hbar N/(m^*\ell)$ and 
$p=4\tau\langle L_B \rangle\ell/h$.
We note that $I_{\rm odd}(p)=I_{\rm odd}(p+4)$ and 
$I_{\rm even}(p)=I_{\rm even}(p+4)$. 
The periodicities of $I_{\rm odd}$ and $I_{\rm even}$ stem from the 
fact that only the states 
$\displaystyle |k_{\alpha}|\leq \sqrt{2m^* E_F}/\hbar$ 
contribute to the current; if $|k_{\alpha}|$ for a given $\alpha$ 
exceeds $\displaystyle \sqrt{2m^* E_F}/\hbar$ by imposing a sufficiently 
large (or small) $\langle L_B \rangle$, 
the state $k_{\alpha}$ becomes vacant and instead the state 
$k_{\alpha}-2\pi/\ell$ is occupied (See reference \cite{persistentderivation} 
for details).

Since precise control of $N$ is difficult experimentally,
we assume an ensemble average over many experimental realizations of 
isolated twisted rings to obtain $(I_{\rm odd}+I_{\rm even})/2$, namely, 
\begin{eqnarray}
I=I(p)=\left\{ 
\begin{array}{ccc}
0 \hspace{2.0cm} &{\rm for}& \hspace{0.5cm} p=0, \\
\displaystyle 
\frac{ev_F}{\ell}(1-p) \hspace{1.0cm} &{\rm for}& \hspace{0.5cm} 0< p <2,
\end{array} 
\right.
\label{eq:matrix}
\end{eqnarray}
where $I(p)=I(p+2)$. 
\section{ESTIMATION OF THE INDUCED CURRENT}
In order to estimate the magnitude of $I$ observed in experiments,
we consider a twisted silver quantum ring.
Successful syntheses of ultrathin crystalline silver nanowires of 
nanometer scale width and micrometer scale length 
have been reported \cite{silverwire1,silverwire3,silverwire4}, 
followed by theoretical studies on their structural 
and transport properties \cite{transport1,transport2,transport3,transport4}.
Such nanowires with high aspect ratios (i.e. the ratio of 
length to width) may be candidates for fabricating a twisted quantum ring.
It should be borne in mind, however, that the applicability of 
our theory is not limited to a specific 
material but to general mesoscopic rings with twisted geometries.

\begin{figure}[t]
\begin{center}
\includegraphics[width=7.5cm]{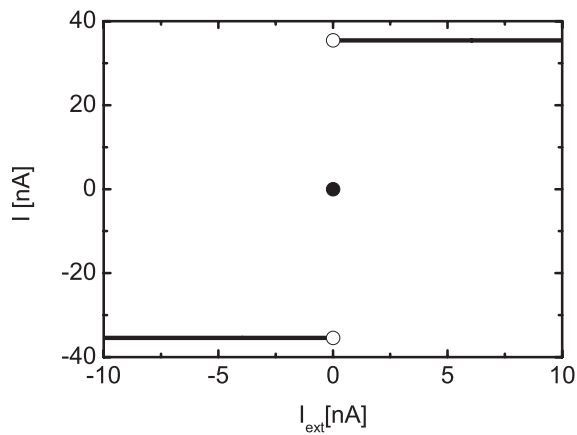}
\caption{\label{fig:fig4} Stepwise behavior of $I$ for the 
twisted ring with $R_1=1.0\mu$m, $R_2=1.0$nm and $\tau=1/\ell$. 
Except at $I_{\rm ext}=0$,
the magnitude of $I$ is almost invariant to the changes in $I_{\rm ext}$
and $\tau$.
}
\end{center}
\end{figure}

Figure \ref{fig:fig4} is a plot of $I$ as a function of 
$I_{\rm ext}$ as given in equation(\ref{eq:matrix}). 
We have set $R_1=1 {\rm \mu}$m, $R_2=1$nm by referring to an actual 
length and radius of the silver nanowires presented in references \cite{silverwire1,silverwire3,silverwire4}, and $\tau = 1/\ell$ 
({\it i.e.,} one twist for one round) for simplicity. 
The Fermi velocity in silver is $v_F=1.39 \times 10^6$m/s \cite{fermivelocity}
, and the characteristic energy scale $\hbar \omega_p$ that 
corresponds to the cross-sectional radius $R_2=1$nm is estimated by 
$\hbar \omega_p = 0.1$ eV from the relation 
$\hbar \omega_p \sim m^*\omega_p^2R_2^2/2$ and 
$m^* = 9 \times 10^{-31}$ kg for silver. 
In figure \ref{fig:fig4}, we observe a stepwise increase 
in $I$ that jumps from $I=-35.4$nA (for $I_{\rm ext} <0$) to 
$I=+35.4$nA (for $I_{\rm ext}>0$). 
Except at $I_{\rm ext}=0$,
the magnitude of $I$ is almost invariant to the changes in $I_{\rm ext}$
and $\tau$.
This constant behavior of $I$ is attributed to the fact that 
under the present conditions, $p$ is much less than unity; 
as a result, $\displaystyle I\sim \frac{ev_F}{\ell}$ for $I_{\rm ext}>0$ and 
$\displaystyle I\sim -\frac{ev_F}{\ell}$ for $I_{\rm ext}<0$, respectively, 
as seen from equation(\ref{eq:matrix}).

The most important observation is the amplitude of $I$ being $35.4$nA 
that is comparable with
the values obtained by using conventional measurement techniques
\cite{persistent1,persistent2,persistent3,persistent4,persistent5,persistent6,persistent7}.
This result indicates the physical significance of the torsion-induced 
quantum phase shift in actual nanostructures with twisted geometries. 
We emphasize that the mechanism by which a persistent current is induced 
in our system differs inherently from its counterpart in an 
untwisted ring, in the latter of which quantum phase shift occurs as the 
result of the application of an external magnetic field
that threads the center of the ring. 
 
\section{CONCLUDING REMARKS}

It deserves comments on other possible apparatus that 
exhibit torsion-induced current flow. 
In the present work, an external current $I_{\rm ext}$ was assumed to 
thread the center of the ring in order to obtain a non-zero 
expectation value of the angular momentum 
of the cross-sectional wave function. 
Differing from the manner, we may directly apply an external magnetic field in 
a direction {\it tangential} to a twisted structure. 
For instance, let us consider a twisted wire (not ring)
both ends of which are connected by a lead, and apply 
a magnetic field of the order of one gauss in a direction 
tangential to the wire. 
Such an apparatus functions in a way similar to that considered in 
Section 3, and therefore, it causes 
torsion-induced current flow in the loop composed of the wire and lead. 
To date, many attempts have been done to synthesize \cite{nature,prl} and 
simulate \cite{prl2,prb} a various kind of twisted nanowires. 
Their results may give a clue to build a set-up toward experimental 
test of our theoretical predictions.

In conclusion, we have demonstrated that a novel type of persistent
current is induced in a quantum coherent ring formed by a long thin 
twisted quantum ring. 
This persistent current is a result of the
geometric torsion of the ring
 that causes a quantum phase shift 
in the eigenstates of the electrons moving in the ring. 
The magnitude of the persistent current is within a realm of 
the results obtained from laboratory experiments; 
this indicates the importance of torsion-induced phenomena in 
influencing the physical properties of actual nanostructures with 
twisted geometries.
\ack

We are grateful to K Yakubo for fruitful discussions, 
and to an anonymous referee for bringing reference \cite{entin}
to our attention. 
One of the authors (HT) adcknowledges K A Mitchell for his helpful 
advices and K W Yu for hospitalities during the stay in The Chinese 
University of Hong Kong. 
HT also thanks the financial support 
from JSPS Research Fellowships for Young Scientists.
HS thanks M Arroyo for his help and hospitality in using the
facility of UPC.
This work is supported by a Grant-in-Aid for 
Scientific Research from the MEXT, Japan.




\section*{References}
\begin{thebibliography}{80}
\bibitem{math1} Jensen H and Koppe H 1971 
{\it Ann. Phys.} \textbf{63} 586

\bibitem{math2}
da Costa R C T 1981 {\it Phys. Rev. A} \textbf{23} 1982

\bibitem{maraner}
Maraner P 1995
{\it J. Phys. A: Math. Gen.} \textbf{28} 2939

%
%
%
\bibitem{math6}
See, for instance, Schuster P C and Jaffe R L 2003
{\it Ann. Phys.} \textbf{307} 132,
and references therein.
%

\bibitem{phys2}
${\rm \acute{E}}$ntin M V and Magarill L I 2001
{\it Phys. Rev.} B \textbf{64} 085330;
${\rm \acute{E}}$ntin M V and Magarill L I 2002 
{\it Phys. Rev.} B \textbf{66} 205308

\bibitem{phys3}
Aoki H, Koshino M, Takeda D, Morise H and Kuroki K 2001
{\it Phys. Rev.} B \textbf{65} 035102

%
%
%

\bibitem{phys8}
Marchi A, Reggiani S, Rudan M and Bertoni A 2005
{\it Phys. Rev.} B \textbf{72} 035403

\bibitem{phys9}
Gravesen J and Willatzen M 2005
{\it Phys. Rev.} A \textbf{72} 032108

%
%
%
%
%
\bibitem{encinosa}
Encinosa M 2006
{\it Phys. Rev.} A \textbf{73} 012102


\bibitem{phys10}
Taira H and Shima H 2007
{\it Surf. Sci.} \textbf{601} 5270


\bibitem{phys12}
Ferrari G and Cuoghi G 2008
{\it Phys. Rev. Lett.} \textbf{100} 230403

\bibitem{phys13}
Atanasov V, Dandoloff R and Saxena A 2009
{\it Phys. Rev.} B \textbf{79} 033404

\bibitem{phys14}
Cuoghi G, Ferrari G and Bertoni A 2009
{\it Phys. Rev.} B \textbf{79} 073410

\bibitem{phys15}
Shima H, Yoshioka H and Onoe J 2009
{\it Phys. Rev.} B \textbf{79} 201401(R); 2010 {\it Physica} E 
doi: 10.1016/j.physe.2009.10.030
  
\bibitem{phys16}
Ono S and Shima H 2009
{\it Phys. Rev.} B \textbf{79} 235407; 2010 {\it Physica} E 
doi: 10.1016/j.physe.2009.11.103


\bibitem{exp1}
Shea H R, Martel R and Avouris P 2000
{\it Phys. Rev. Lett.} \textbf{84} 4441

\bibitem{exp2}
Schmidt O G and Eberl K 2001
{\it Nature} \textbf{410} 168


\bibitem{exp4}
Lorke A, B\"ohm S and Wegscheider W 2003
{\it Superlatt. Microstr.} \textbf{33} 347
  
  
\bibitem{exp6}
Onoe J, Nakayama T, Aono M and Hara T 2003
{\it Appl. Phys. Lett.} \textbf{82} 595

\bibitem{jpcm}
McIlroy D N, Alkhateeb A, Zhang D, Aston D E,
Marcy A C and Norton M G 2004
{\it J. Phys.: Cond. Matt.} \textbf{16} R415


\bibitem{exp8}
Gupta S and Saxena A 2009
{\it J. Raman. Spectrosc.} \textbf{40} 1127


\bibitem{torsion1}
Takagi S and Tanzawa T 1992
{\it Prog. Theor. Phys.} \textbf{87} 561
  
\bibitem{torsion2}
Mitchell K A 2001
{\it Phys. Rev.} A \textbf{63} 042112
  
\bibitem{entin}
Magarill L I and ${\rm \acute{E}}$ntin M V 2003
{\it J. Exp. Theor. Phys.} \textbf{96} 766

\bibitem{persistent1}
L\'evy L P, Dolan G, Dunsmuir J and Bouchiat H 1990
{\it Phys. Rev. Lett.} \textbf{64} 2074
  
\bibitem{persistent2}
Chandrasekhar V, Webb R A, Brady M J, Ketchen M B, Gallagher W J
and Kleinsasser A 1991
{\it Phys. Rev. Lett.} \textbf{67} 3578
  
\bibitem{persistent3}
Mailly D, Chapelier C, and Benoit A 1993
{\it Phys. Rev. Lett.} \textbf{70} 2020

\bibitem{persistent4}
Fuhrer A, L$\rm \ddot{u}$scher S, Ihn T, Heinzel T, Ensslin K,
Wegscheider W and Bichler M 2001
{\it Nature} \textbf{413} 822
  
\bibitem{persistent5}
Kleemans N A J M, Bominaar-Silkens I M A, Fomin V M,
Gladilin V N, Granados D, Taboada A G, Garcia J M, Offermans P,
Zeitler U, Christianen P C M, Maan J C, Devreese J T and Koenraad P M 2007
{\it Phys. Rev. Lett.} \textbf{99} 146808

\bibitem{persistent6}
Bluhm H, Koshnick N C, Bert J A, Huber M E and Moler K A 2009
{\it Phys. Rev. Lett.} \textbf{102} 136802

\bibitem{persistent7}
Bleszynski-Jayich A C, Shanks W E, Peaudecerf V, Ginossar E, von Oppen F, 
Glazman L, and Harris J G E 2009
{\it Science} \textbf{326} 272

\bibitem{persistentderivation2}
Imry Y 2002
{\it Introduction to Mesoscopic Physics}
(Oxford University Press)

\bibitem{shima}
Shima H and Nakayama T 2010 
{\it Higher Mathematics for Physics and Engineering}
(Berlin: Springer-Verlag)


\bibitem{Fock}
Fock V 1928
{\it Z. Phys.} \textbf{47} 446

\bibitem{Darwin}
Darwin C G 1930
{\it Proc. Camb. Philos. Soc.} \textbf{27} 86


\bibitem{persistentderivation}
Cheung H F, Gefen Y, Riedel E K and Shih W H 1988
{\it Phys. Rev.} B \textbf{37} 6050

\bibitem{imry}
Bary-Soroker H, Entin-Wohlman O and Imry Y 2008
{\it Phys. Rev. Lett.} \textbf{101} 057001

\bibitem{silverwire1}
Hong B H, Bae S H, Lee C W, Jeong S and Kim K S 2001
{\it Science} \textbf{294} 348


\bibitem{silverwire3}
Sun Y, Mayers B, Herricks T and Xia Y 2003
{\it Nano. Lett.} \textbf{3} 955

\bibitem{silverwire4}
Graff A, Wagner D, Ditlbacher H and Kreibig U 2005
{\it Eur. Phys. J.} D \textbf{34} 263

\bibitem{transport1}
Rodrigues V, Bettini J, Rocha A R, Rego L G C and Ugarte D 2002
{\it Phys. Rev.} B \textbf{65} 153402

\bibitem{transport2}
Zhao J, Buia C, Han J and Lu J P 2003
{\it Nanotechnology} \textbf{14} 501

\bibitem{transport3}
Elizondo S L and Mintmire J W 2006
{\it Phys. Rev.} B \textbf{73} 045431

\bibitem{transport4}
Jia J, Shi D, Zhao J and Wang B 2007
{\it Phys. Rev.} B \textbf{76} 165420

\bibitem{fermivelocity}
Mitchell J W and Goodrich R G 1985
{\it Phys. Rev.} B \textbf{32} 4969



%
%
%
%
%
%
%
%
%
%
%

\bibitem{nature} 
Cohen-Karni T, Segev L, Srur-Lavi O, Cohen S R and Joselevich E 2006
{\it Nature Nanotech.}  \textbf{1} 36

\bibitem{prl} 
Nagapriya K S, Goldbart O, Kaplan-Ashiri I, Seifert G, Tenne R and Joselevich E
 2008
{\it Phys. Rev. Lett.}  \textbf{101} 195501

\bibitem{prl2}
Arias I and Arroyo M 2008
{\it Phys. Rev. Lett.}  \textbf{100} 085503

\bibitem{prb} 
Wang Z, Zu X, Gao F and Weber W J 2008
{\it Phys. Rev.} B  \textbf{77} 224113


\end{thebibliography}
\end{document}